\documentclass[12pt]{iopart}
\usepackage[latin1]{inputenc}
\usepackage{graphicx}
\usepackage{dcolumn}
\usepackage{bm}
\begin{document}
    \title[Biot-Savart inversion in thick superconductors]{Significance of solutions of the inverse Biot-Savart problem in thick superconductors}
\author{M.~Eisterer}
\address{Atominstitut der Österreichischen Universitäten, A-1020 Vienna, Austria}

\ead{eisterer@ati.ac.at}


\begin{abstract}
The evaluation of current distributions in thick superconductors
from field profiles near the sample surface is investigated
theoretically. A simple model of a cylindrical sample, in which
only circular currents are flowing, reduces the inversion to a
linear least squares problem, which is analyzed by singular value
decomposition. Without additional assumptions about the current
distribution (e.g. constant current over the sample thickness),
the condition of the problem is very bad, leading to unrealistic
results. However, any additional assumption strongly influences
the solution and thus renders the solutions again questionable.
These difficulties are unfortunately inherent to the inverse
Biot-Savart problem in thick superconductors and cannot be avoided
by any models or algorithms.
\end{abstract}

\maketitle

\section{Introduction}

The calculation of two dimensional current distributions from the
magnetic field generated by them was successfully performed with
various algorithms
\cite{Rot89,Bra92,Gra94,Nic96,Wij96,Pas97,Joo98}. All these
algorithms are based on the assumption, that the current flows in
infinitely thin layers \cite{Rot89,Gra94,Pas97} or that the
current is uniform \cite{Wij96,Joo98} (or averaged \cite{Bra92})
over the sample thickness. In most cases, only the z-component of
the magnetic field (orthogonal to the sample surface) is measured
on a 2D grid. From this discrete field distribution the two
components (x,y) of the current are calculated on a similar grid
within the sample by matrix inversion \cite{Gra94,Nic96,Wij96} or
by fast Fourier transformation \cite{Rot89,Pas97,Joo98}. It is
straightforward to extend these algorithms to thick
superconductors, if the current can be assumed to be constant over
the sample thickness \cite{Wij98,Car03}. Unfortunately, this
condition is never fulfilled in real superconuctors due to
material inhomogeneities and due to the self field. The influence
of the inhomogeneities can easily be seen by the differences
between the remnant field profiles taken at the top and the bottom
surface of thick bulk samples \cite{Gon03}. Although the effect of
the self field cannot be detected directly, it is obvious that the
changing magnitude (largest in central layers) and direction of
the self field results in a z-dependence of the current density.
Nevertheless, solutions obtained under these wrong assumptions are
used for the analysis of spatial variations of the sample
properties. The aim of this paper is to point out that it is in
principle not possible to calculate the current distribution
within a thick superconductor from the z-component of the remnant
field profile, if a realistic experimental error is taken into
account. It will also be shown, that any unjustified assumptions
lead to artifacts in the resulting solutions. The influence of the
experimental error, of noise reduction and of additional (wrong)
assumptions is demonstrated for a simplified model system and the
resulting solutions are compared with the true (known) current
distribution.

\section{Model System}

\begin{figure} \centering \includegraphics[clip,width
=0.25\columnwidth]{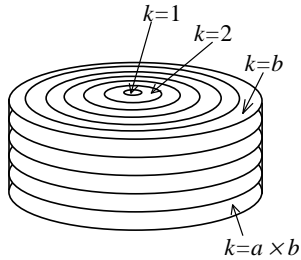} \caption{Radial and axial
division of the mathematical sample into toroids.}
\label{division}
\end{figure}

The sample is chosen to be a cylinder with radius $R$ and
thickness $d$. This sample is divided $a$ times axially
(perpendicular to the sample axis), resulting in $a$ layers. Each
layer is divided again $b$ times radially (into hollow cylinders),
leading to $n:=a\times b$ toroids of rectangular cross section
(Fig. \ref{division}).
The toroids are labeled by $k$, starting with the innermost toroid
of the first layer. The numeration continues in the first layer
until $k=b$, then jumps to the innermost toroid of the second
layer and so on. Currents are assumed to flow circularly with a
constant current density in each toroid. The magnetic induction at
a point $\overrightarrow{r}$, generated by each toroid, is then
proportional to the actual current density in the toroid. This
leads to a system of linear equations between the current
densities in each toroid $j_{k}$ and the z-component of the
magnetic induction at $m$ discrete points
$B_{z}(\overrightarrow{r_{l}})$ (the data points $B_{l}$):
\begin{equation}
B_{l}=\sum_{k=1}^{n}M_{l,k}j_{k} \hspace{9 mm}  l=1....m
\end{equation}
or more compactly
\begin{equation}
\label{leqs}
 \overrightarrow{B}=M\overrightarrow{j}.
\end{equation}
The m-dimensional vector \overrightarrow{B} and the n-dimensional
vector \overrightarrow{j} are representing discrete values of
$B_{z}$ at $m$ points outside the sample and of the current
densities in the $n$ toroids, respectively. The coefficients
$M_{l,k}$ are calculated by integrating the Biot-Savart law over
the volume of the corresponding toroid. In the following it is
assumed that the number of data points $m$ exceeds the number of
toroids $n$, i.e. the equation is overdetermined, which is
favorable for the suppression of experimental noise.

This model system was chosen because of its simplicity. It is
radially symmetric, which reduces the number of data points
significantly. At the same spatial resolution the field profile is
determined by $m$ instead of $4m^{2}$ data points, e.g., for a
grid width of 0.5 $mm$, $m$ is 29 for a field profile from $r=0$
to $r=14 \, mm$ instead of $4m^{2}=3364$ for a two dimensional
quadratic grid. A similar reduction is obtained for the grid
inside the sample, but in this case the knowledge of the current
direction leads to an additional reduction of the number of
unknown parameters by a factor of three (magnitude of the current
density instead of its three components) and current conservation
is fulfilled automatically. The resulting systems of equations are
relatively small and can be solved numerically without any
additional algorithmic problems, which would be inherent to large
systems of equations. The condition of this simplified problem is
certainly better than in a more complex system. It is possible
with this model to study the inverse Biot-Savart problem without
any numerical problems and it is obvious that a more realistic
model behaves worse. The model restricts the possible solutions to
circular currents, which can also be present in real samples.
Therefore, any algorithm for the inversion of more general current
distributions should also be able to invert circular currents.
Thus, the results obtained from this model system are also valid
for any other algorithmic implementation of the inverse
Biot-Savart problem in thick superconductors.

\section{Condition Number}

A first insight into the problem can be obtained by calculating
the condition number $K$ of the coefficient Matrix $M$ \cite{CoN}.
The condition number reflects the error propagation:
\begin{equation}
\frac{\Delta B}{B} \leq K\frac{\Delta j}{j}
\label{equk}
\end{equation}
The relative error of the calculated current density can be $K$
times larger than the relative experimental error of the magnetic
field values. In principle, significant results can only be
obtained, if the product of the condition number and the relative
experimental error is much smaller than one. Since equality is the
worst case scenario of the inequality \ref{equk}, reasonable
results can usually be obtained as long as this product is not
much larger than unity, but in this case a careful analysis of the
error propagation is needed.

\begin{figure} \centering \includegraphics[clip,width
= 0.5\columnwidth]{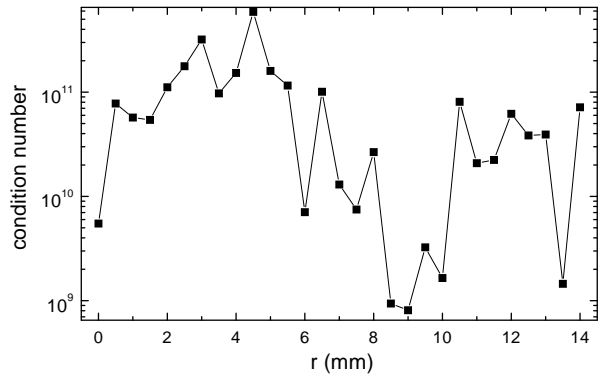} \caption{Condition number in the case
of an axial field profile as a function of the radial coordinate.}
\label{cond}
\end{figure}

The condition number is reasonably small, if the sample is divided
only radially (a=1). Assuming a sample of radius $R=12.5 \, mm$
with a height of $d=500 \, nm$ and inverting a radial flux profile
with 29 points between $r=0$ and $r=14 \, mm$, measured $500 \,
nm$ above the sample surface, the condition number is 5.7 for ten
radial divisions (b=10). This small thickness and the small gap
between the sample and the field profile is representative for
magnetooptical measurements on thin films, where inversion schemes
are well established \cite{Pas97,Joo98}. This quite small
condition number (K cannot be smaller than one) becomes 16.3, if
the sample thickness is enhanced to $d=10 \, mm$, which is typical
for melt textured bulk samples. If a gap of $\Delta z=0.2 \, mm$
is assumed (representative for Hall probe measurements) $K$ is
further increased to 27.3. These values are still suitable for a
proper inversion. Therefore, numerically stable \cite{Wij98,Car03}
(but not necessarily correct !) solutions can be obtained from the
inversion of the Biot-Savart law, if the current is assumed to be
homogeneous along the sample thickness. The situation immediately
changes without this restriction. If the sample is divided only
axially (a=10, b=1), the condition number for the same number of
free parameters and for the same geometry becomes
4.7$\times$$10^{8}$. One cannot expect to get correct results in
this case for any reasonable experimental error bars. In order to
exclude the possibility, that this awful condition is a
consequence of the {\it radial} flux profile for the calculation
of an {\it axial} current distribution, the condition number was
calculated for axial field profiles (29 points from $\Delta z=0.2$
to $\Delta z=28.2 mm$) at different radial positions (Fig.
\ref{cond}).
The condition number depends on the radial position of the axial
field profile, but does not become smaller than 8.1$\times$$
10^{8}$. Providing a radial field profile at each $\Delta z$ (841
data points) reduces $K$ only slightly to 1.4$\times$$10^{8}$. It
is obviously impossible to calculate the current densities of only
10 layers from 841 data points (with realistic experimental
errors). This result indicates immediately that a general
inversion of the Biot-Savart law is simply not possible for thick
superconductors.

\section{Single Value Decomposition (SVD)}

\begin{figure} \centering \includegraphics[clip,width
= 0.5\columnwidth]{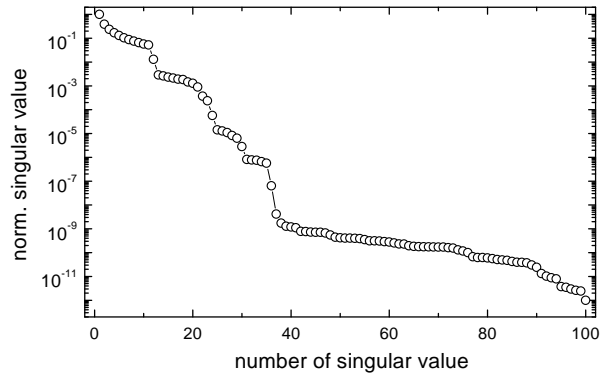} \caption{Singular values of the
model system.} \label{sinval}
\end{figure}

The single value decomposition (SVD) is a powerful tool for the
solution of systems of linear equations \cite{SVD}. It not only
solves the equations in a numerically stable form, but also gives
insights into the problems, which may arise. The $m \times n$
matrix $M$ of Equ. \ref{leqs} can be decomposed into three
matrices: $ M=U \Lambda V^{T} $. Since the $n \times n$ matrix $V$
is orthonormal, its columns $\overrightarrow{v_{i}}$ can be
interpreted as basis vectors for the representation of the current
densities. $\Lambda$ is a $n \times n$ diagonal matrix. The
(positive) diagonal elements are called singular values
$\lambda_{i}$, with
$\lambda_{1}\geq\lambda_{2}\geq...\geq\lambda_{n}$. While it is
generally possible that some singular values become zero, this
case will be excluded in the following, since it does not occur in
the actual model systems. The columns $\overrightarrow{u_{i}}$ of
the $m \times n$ matrix $U$ are pairwise orthonormal and are basis
vectors for the representation of \overrightarrow{B}. Since $m$
was assumed to be larger than $n$ the basis vectors
$\overrightarrow{u_{i}}$ only span a subspace of
\overrightarrow{B}, the range of $M$. $M$ maps each basis vector
$\overrightarrow{v_{i}}$ to $\overrightarrow{u_{i}}$ multiplied by
the corresponding singular value $\lambda_{i}$:
\begin{equation}
Mv_{i}=\lambda_{i}u_{i}.
 \label{mapM}
\end{equation}
Problems occur, if some singular values are much smaller than
others. Current components corresponding to such small singular
values only add very little to the field profile and are very hard
or impossible to detect, especially if their contribution becomes
smaller than the experimental error. To be more concrete, the
singular values were calculated for a sample, which is
representative for melt textured bulk samples ($R=12.5 \, mm, \,
d=10 \, mm$). It was divided into 100 toroids ($a=b=10$) and 824
data points were assumed to be available at 29 radial flux
profiles with 29 points each ($\Delta z=0.2...28.2 \,mm, \,
r=0...14 \, mm$). The singular values are plotted in Fig.
\ref{sinval}, normalized by the largest singular value
$\lambda_{1}$, since their absolute values depend only on the
actual units.
They vary by twelve orders of magnitude, which explains the bad
condition of the problem. If the experimental error is about 1 \%,
only the first 12 current components $\overrightarrow{v_{i}}$ are
expected to generate a signal that is larger than this
experimental error. In principle, this is only true, if the
magnitude of the current components with large singular values is
not much smaller than the magnitude of the other current
components, but this condition is fulfilled for any reasonable
current distribution. Since most of the current components are not
generating a significant signal, they cannot be determined. On the
other hand, one can at least calculate the current distribution
within the sample projected to this 12 dimensional subspace. The
image of this subspace can be defined as the "experimental range"
of the matrix $M$.

Although an experimental error of 1 \% seems to be high for the
measurement of the magnetic field, the total expected error for a
typical experiment is of that order of magnitude, since other (in
principle not stochastic) sources of error  have to be added. The
gap between the Hall probe or the magnetooptical layer is usually
not known exactly, relaxation of the currents decreases the field
during the measurement, oxygen pick-up in liquid nitrogen can
change its temperature and so on. A decrease of the experimental
error leads to an increase of the dimension of the experimental
range, but the basic problem remains the same.

\begin{figure} \centering \includegraphics[clip,width
= 0.5\columnwidth]{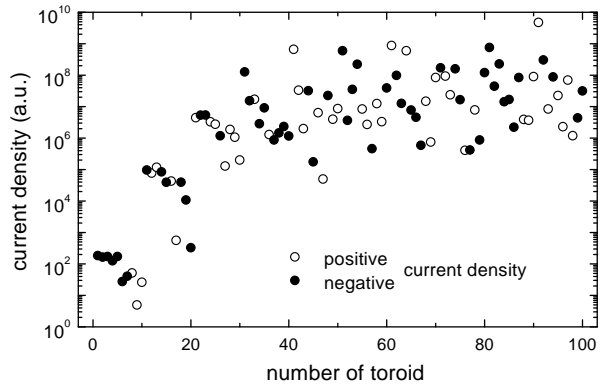} \caption{Best approximating current
distribution for data with typical experimental error.} \label{jn}
\end{figure}

The best approximating solution (in the least square sense) can be
obtained easily after the single value decomposition:
\begin{equation}
\overrightarrow{j}=V\Lambda^{-1}U^{T}\overrightarrow{B}
\label{invers}
\end{equation}
The multiplication with $U^{T}$ projects \overrightarrow{B} to the
range of M and performs a basis transformation (new basis:
$u_{i}$). All experimental errors, which are out of the range of
M, are mapped to zero and do not induce any error in the
calculated currents at all. The coefficients (in the new
representation) of the remaining \overrightarrow{B} are then
divided by the corresponding singular values $\lambda_{i}$.
Finally, V performs another basis transformation. Components of
the measured magnetic induction pointing in the direction
$\overrightarrow{u_{i}}$ are amplified by $1/\lambda_{i}$.
Especially components of the experimental error corresponding to a
small singular value (large $i$) will be strongly amplified. From
the following worst case scenario, the condition number of the
system can be easily calculated. The current distribution
$\overrightarrow{j}$ within the sample is just proportional to
$\overrightarrow{v_{1}}$ generating a field $\overrightarrow{B}=|
\overrightarrow{j} | \lambda_{1}\overrightarrow{u_{1}}$. An
experimental error $\overrightarrow{\Delta B}$ proportional to
$\overrightarrow{u_{n}}$ induces an error in the derived current
distribution $\overrightarrow{\Delta j}=| \overrightarrow{\Delta
B} | /\lambda_{n}\overrightarrow{v_{n}}$. This leads immediately
to the inequation \ref{equk} with
$K=\frac{\lambda_{n}}{\lambda_{1}}$. In the system under
consideration the condition number is about $10^{12}$ (Fig.
\ref{sinval}). In order to illustrate the influence of this awful
condition number, the field profile $\overrightarrow{B_{0}}$
generated by a totally homogeneous current distribution of one (in
arbitrary units) was calculated, i.e. the same current density is
assumed to flow in each toroid. To check for numerical problems,
$\overrightarrow{B_{0}}$ was inverted with relation \ref{invers}.
The maximum deviation from unity due to round-off errors was as
small as 0.14 \%, i.e. there are no numerical problems. Each
component of $\overrightarrow{B_{0}}$ was then multiplied by a
random number between 0.99 and 1.01, which simulates a
(stochastic) experimental error of 1 \%. The inversion leads to a
completely different current distribution (Fig. \ref{jn}) with
huge (positive and negative) current densities.
The largest value is 4.7$\times$$10^{9}$, about half of the upper
limit for the error propagation $K \Delta B=10^{10}$. With the aid
of the single value decomposition, not only the condition number
of the whole problem can be derived, but the condition number of
each component of the field (coefficients of $u_{i}$) or of the
corresponding current density ($v_{i}$) can be obtained
separately. As pointed out above, only the first twelve components
are significant (if an experimental error of 1 \% is assumed),
i.e. they are expected to generate a signal that is larger than
the experimental error. Inverting only these twelve significant
field components of $\overrightarrow{B_{0}}$ (or equivalently,
projecting the whole solution to the 12 dimensional subspace of
$\overrightarrow{j}$) leads to a current distribution, which is
far from being constant and is in average, much smaller than one
(0.73).
\begin{figure} \centering \includegraphics[clip,width
= 0.5\columnwidth]{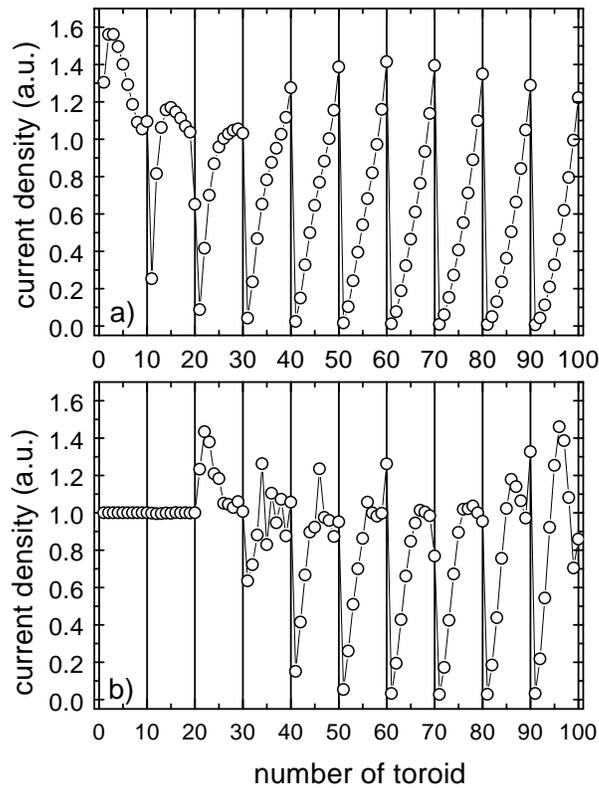} \caption{Inversion of the
"experimental range" for an assumed experimental error of (a)
$10^{-2}$ and (b) $10^{-9}$.} \label{j12/50}
\end{figure}
In Fig. \ref{j12/50}a, data points between two grid lines
represent the radial current distribution within one layer. This
obviously wrong current distribution fits the data perfectly, even
better than the assumed error of 1 \% in the Euclidean norm. In
this norm the deviation is only 4.5$\times$$10^{-4}$, in the
maximum norm about $10^{-3}$. Calculating the relative deviation
in each point separately, it is found to be always smaller than
3.5$\times$$ 10^{-3}$ (in 11 points out of 841 this cannot be
done, because the value is numerically zero there). This is a
general behavior of ill conditioned problems. The solution nicely
agrees with the data, as long as it is correct in the significant
subspace. In the remaining space the solution can be assumed more
or less arbitrarily. Therefore, agreement with the inverted data,
does not indicate the correctness of the solution. If the noisy
version of $\overrightarrow{B_{0}}$ is inverted in that way, the
"solution" looks quite similar, with deviations of up to 14 \%,
significantly lower than the theoretical upper limit of the error
propagation (100 \%). This justifies the rule of thumb that the
experimental error times the condition number should not be much
larger than one. The projection to the significant subspace
represents some sort of filtering. {\it Any} filtering (e.g.,
disregarding the highest frequency of the Fourier transformation)
not only reduces the noise, but also changes the solution, because
it cannot distinguish between experimental noise and the "true"
signal. Its influence on the solution is the stronger the worse
the condition of the problem and one should be very careful with
solutions obtained after filtering. In the present case the
decrease of the currents from the top to the bottom could be
interpreted as a deterioration of the sample properties with
increasing distance from the seed, the higher currents at the
sample edges as the influence of the self field. Both explanations
are expected and, therefore, plausible, but {\it cannot} be
derived from the present data, since the field profile was
calculated assuming a constant current density.

Figure \ref{j12/50}b shows the projection of the exact solution to
the first 50 $v_{i}$s, which would require the experimental error
to be smaller than $10^{-9}$. Even in this unrealistic case the
correct solution is obtained only in the uppermost two layers, the
situation in the bottom layers does not improve significantly.
Although it might be possible to improve the condition of this
specific problem by a few orders of magnitude by optimizing the
grid and the sample division, this would not be sufficient, even
if the experimental error could be reduced by one or two orders of
magnitude. In a more realistic system, i.e. allowing the currents
to flow in arbitrary directions, the condition is expected to be
even worse.

\section{Homogeneous Current along the Sample Thickness}

\begin{figure} \centering \includegraphics[clip,width
= 0.5\columnwidth]{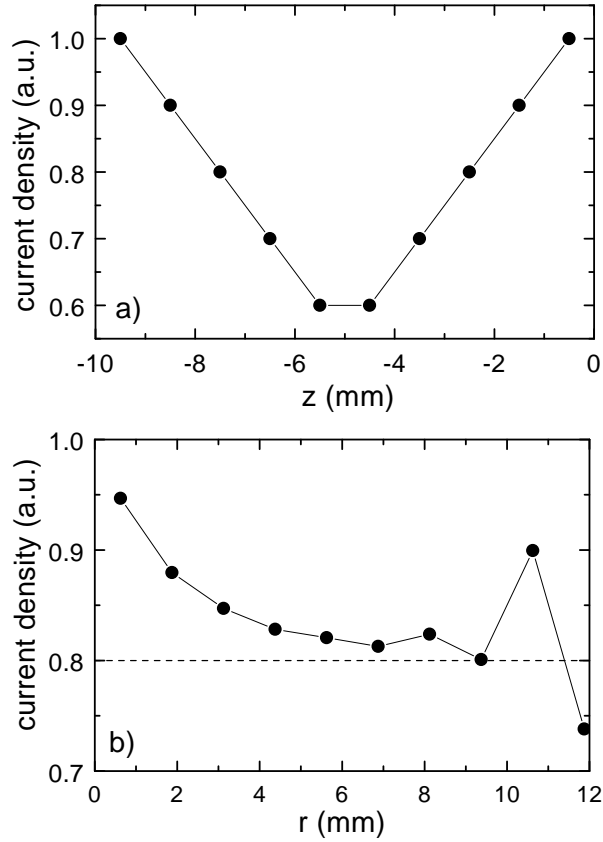} \caption{(a) Assumed axial current
distribution and (b) calculated radial current distribution. The
correct average current is constant 0.8 (broken line).}
\label{jcz/r}
\end{figure}

As already pointed out, one possibility to obtain a reasonable
condition for the inverse Biot-Savart problem is to assume the
currents to be homogeneous along the sample thickness
\cite{Wij98,Car03}. The influence of a violation of this condition
is discussed in this section. The field profile of a a sample
($R=12.5 \, mm, d=10 \, mm$) was calculated for a radially
constant but axially changing current density ($a=10,\, b=1$). The
model current distribution is plotted in figure \ref{jcz/r}a,
$z=0$ and $z=-10 \, mm$ correspond to the top and bottom surface,
respectively.
Such a behavior is expected even in a totally homogeneous sample
due to the self field, which is largest in the central layers.
Since the critical current density decreases with field (at least
at low fields), the currents are largest at the surface layers.
The calculated field profile was then inverted (wrongly) assuming
the current to be constant over the sample thickness ($a=1, \,
b=10$). The resulting radial current distribution (Fig.
\ref{jcz/r}b) is larger than the correct value of 0.8 averaged
over $z$ (except near the sample edge) and a pronounced radial
dependence is observed. The assumption of homogeneous current
along the sample thickness, which is never fulfilled in thick
superconductors, is not appropriate for the inverse Biot-Savart
problem in these samples, since neither the correct magnitude nor
the correct radial dependence of the current are obtained.

\section{Conclusions}

It was shown that it is generally not possible to derive the three
dimensional current distribution within a thick superconductor
from a three dimensional distribution of the z-component of the
magnetic induction, which is generated by this current
distribution. Under additional assumptions or by filtering,
reasonably looking but wrong current distributions are obtained.
The field generated by these wrong current distributions agrees
with the correct field within experimental accuracy. Agreement
between measured data and calculated current densities is,
therefore, no indication for the correctness of the solution or of
any of the underlying assumptions. Such solutions are most
probably wrong and may lead to wrong conclusions.


\section*{References}
\begin{thebibliography}{12}

\bibitem{Rot89} Roth  B J, Sepulveda N G and Wikswo J P 1989 {\it J. Appl. Phys.} {\bf 65} 361
\bibitem{Bra92} Brandt E H 1992 {\it Phys. Rev.} B {\bf 46} 8628
\bibitem{Gra94} Grant P D, Denhoff M W, Xing W, Brown P, Govorkov S, Irwin J C, Heinrich B, Zhou H, Fife A A and Cragg A R
1994 {\it Physica} C {\bf 229} 289
\bibitem{Nic96} Niculescu H, Saenz A, Khankhasayev M and Gielisse P J 1996 {\it Physica} C {\bf 261} 12
\bibitem{Wij96} Wijngaarden R J, Spoelder H J W, Surdeanu R and Griessen R 1996 {\it Phys. Rev.} B {\bf 54} 6742
\bibitem{Pas97} Pashitski A E, Gurevich A, Polyanskii A A, Larbalestier D C, Goyal A, Specht E D, Kroeger D M, DeLuca J A and
Tkaczyk J E 1997 {\it Science} {\bf 275} 367
\bibitem{Joo98} Jooss Ch, Warthmann R, Forkl A and Kronmüller H 1998 {\it Physica} C {\bf 299} 215
\bibitem{Wij98} Wijngaarden R J, Heeck K, Spoelder H J W, Surdeanu R and
Griessen R 1998 {\it Physica} C {\bf 295} 177
\bibitem{Car03}  Carrera M, Amorós J, Carrillo A E, Obradors X
and Fontcuberta J 2003 {\it Physica} C {\bf 385} 539
\bibitem{Gon03} Gonzalez-Arrabal R, Eisterer M
and Weber H W 2003 {\it J. Appl. Phys.} {\bf 93} 4734
\bibitem{CoN} A function for the calculation of the condition number
is included in numerical software packages for matrix
manipulations.
\bibitem{SVD} Details can be found in textbooks on numerical mathematics.
Algorithms are provided by numerous numerical software packages.

\end{thebibliography}
\end{document}